\documentclass[11pt]{article}
\usepackage{amsmath}
\usepackage{pgf} 

\usepackage{fullpage}

\title{\uppercase{Rotating Fermions}}

\author{\uppercase{Victor E. Ambru\c{s}} and \uppercase{Elizabeth Winstanley}\\
{\scriptsize \em School of Mathematics and Statistics, University of Sheffield, United Kingdom}}

\begin{document}

\maketitle

\begin{abstract}
We investigate the rigidly rotating quantum thermal distribution of fermions
in flat space-time. We find that thermal states diverge on the
speed of light surface. We remove the divergences by enclosing the system
inside a cylindrical boundary and investigate thermal expectation values and
the Casimir effect for two sets of boundary conditions.
\end{abstract}

\section{Introduction}\label{ave:intro}
The authors of Ref.~\cite{ave:duffy_ottewill} have shown that thermal states
are ill-defined for quantum scalar particles on an unbounded, rigidly rotating,
space-time, but are regular if the system is enclosed in a boundary
inside the speed of light surface (SOL).

We perform a similar analysis for fermions. We discover that the
difference between Bose-Einstein and Fermi-Dirac statistics allows fermions
to exist in thermal states which diverge on the SOL, but are perfectly regular
if enclosed inside a boundary located within the SOL.

\section{Rotating fermions on the unbounded Minkowski space-time}\label{ave:unb}
We consider a Minkowski space-time rotating with a constant angular velocity
$\mathbf{\Omega}$. The line element in cylindrical coordinates, with the
$z$ axis pointing along $\mathbf{\Omega}$, can be obtained from the
Minkowski metric by making the substitution
$\varphi \rightarrow \varphi + \Omega t$:
\begin{equation}\label{ave:ds}
 ds^2 = -(1 - \rho^2 \Omega^2) dt^2 + 2\rho^2 \Omega\,dt\,d\varphi + d\rho^2 +
 \rho^2 d\varphi^2 + dz^2.
\end{equation}
We define the surface $\rho\Omega = 1$ as the speed of light surface (SOL),
since on this surface, co-rotating observers move at the speed of light.

The solutions of the Dirac equation on the metric in Eq.~\eqref{ave:ds} can
be written as:
\begin{equation}\label{ave:solU}
 U_{k}(x) = \frac{1}{2\pi} e^{-i \widetilde{E}_k t}
 u_{k}(\rho,\varphi,z),
\end{equation}
where $u_{k}$ is a four-spinor and $k$ is a generic label
distinguishing between independent modes. The eigenvalue
$\widetilde{E}_k$ of the Hamiltonian is related to the Minkowski energy $E_k$
through $\widetilde{E}_k = E_k - \Omega (m + \tfrac{1}{2})$, where
$m+\tfrac{1}{2}$ is the eigenvalue of the angular momentum along the rotation
axis.

If $b_{k}$ is the creation operator for quanta described
by the modes in Eq.~\eqref{ave:solU}, we can construct the
following thermal expectation value (t.e.v.) at inverse temperature
$\beta$ \cite{ave:vilenkin}:
\begin{equation}
 \left<b_k^\dagger b_{k'}\right>_\beta =
 \frac{\delta_{kk'}}{e^{\beta \widetilde{E}_k} + 1}.
\end{equation}
For the vacuum expectation value of products of the form
$b_k^\dagger b_{k'}$ to vanish, the above expression must go to $0$ as
$\beta \rightarrow \infty$. Therefore, we must restrict particle modes to
modes with positive frequency (i.e. positive $\widetilde{E}_k$), as discussed
in Ref.~\cite{ave:iyer}.
Allowing particle modes to have negative frequency
would give rise to temperature-independent terms in t.e.v.'s, similar to those
obtained in Ref.~\cite{ave:vilenkin}.

We now present the t.e.v. of the energy density using the quantisation
proposed by Ref.~\cite{ave:iyer} ($\varepsilon = 1 - \rho^2\Omega^2$):
\begin{equation}\label{ave:Ttt}
 \left<:T_{tt}:\right>_{\beta} =
 \frac{7\pi^2}{60\beta^4\varepsilon} + \frac{\Omega^2}{8\beta^2}
 \left(\frac{4}{3\varepsilon^{2}} - \frac{1}{3\varepsilon}\right).
\end{equation}
Similar expressions can be obtained
for the other non-vanishing components of the stress-energy tensor (i.e. for
$T_{t\varphi}$, $T_{\rho \rho }$,
$T_{\varphi \varphi }$ and $T_{z z}$),
as well as for the
neutrino current parallel to the rotation axis \cite{ave:rotfer}.
A common feature of these
t.e.v.'s is their divergent behaviour as the SOL is
approached, which we illustrate by the thin curves in Fig.~\ref{ave:plots}.

For the scalar field, the indefinite norm requires particles to be
described by modes of positive norm \cite{ave:letaw_pfautsch}, forcing
negative frequency modes into the set of particle modes. Moreover,
the Bose-Einstein density of states factor diverges for frequencies approaching
$0$, even when the corresponding Minkowski energy is finite, requiring thermal
states to have an infinite energy density at each point in the space-time.
\def\figsubcap#1{\par\noindent\centering\footnotesize(#1)}
\begin{figure}[!ht]
  \begin{minipage}[b]{0.47\linewidth}
    \centering
    \includegraphics[width=\linewidth]{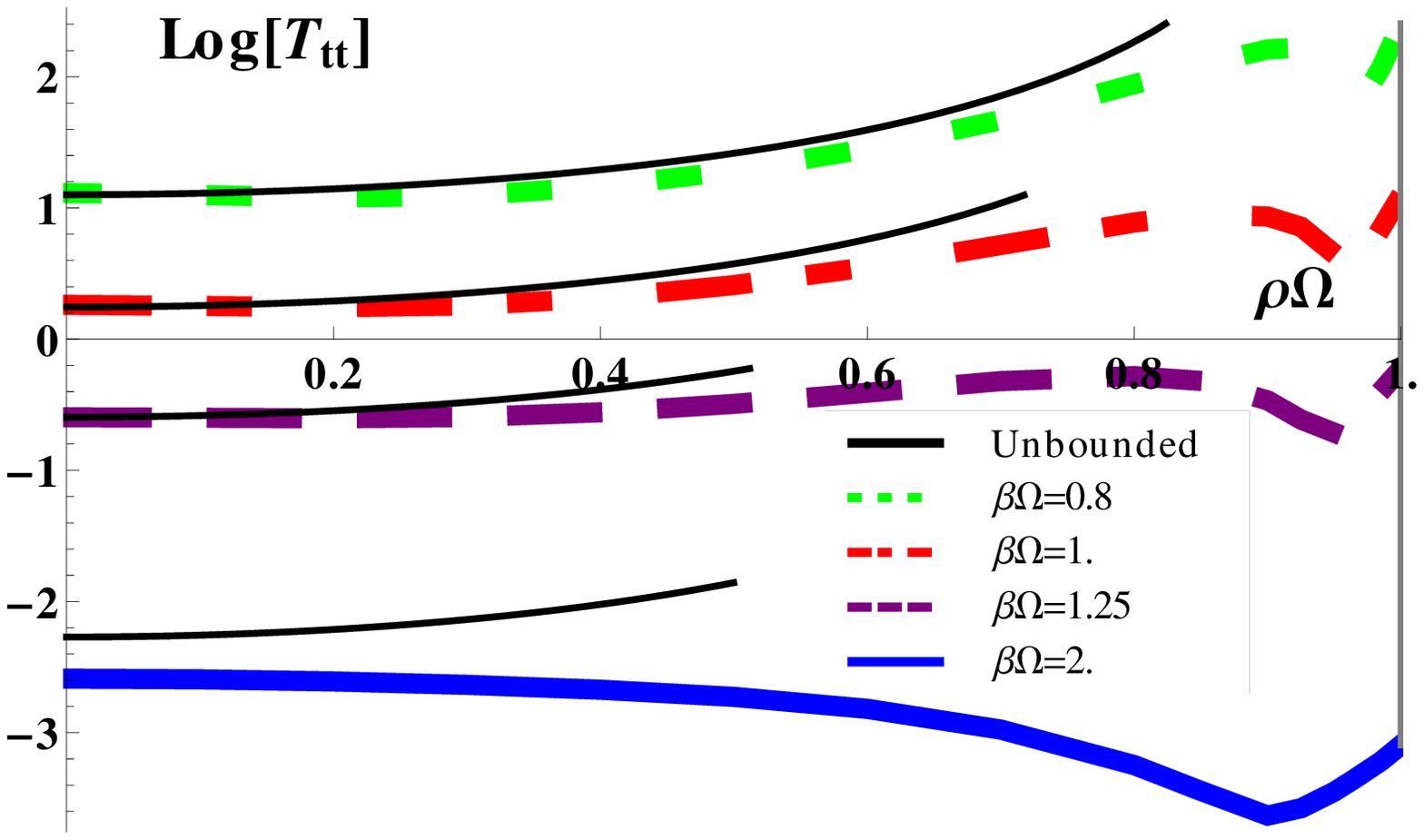}
    \figsubcap{a}
  \end{minipage}
  \hspace{0.03\linewidth}
  \begin{minipage}[b]{0.47\linewidth}
    \centering
    \includegraphics[width=\linewidth]{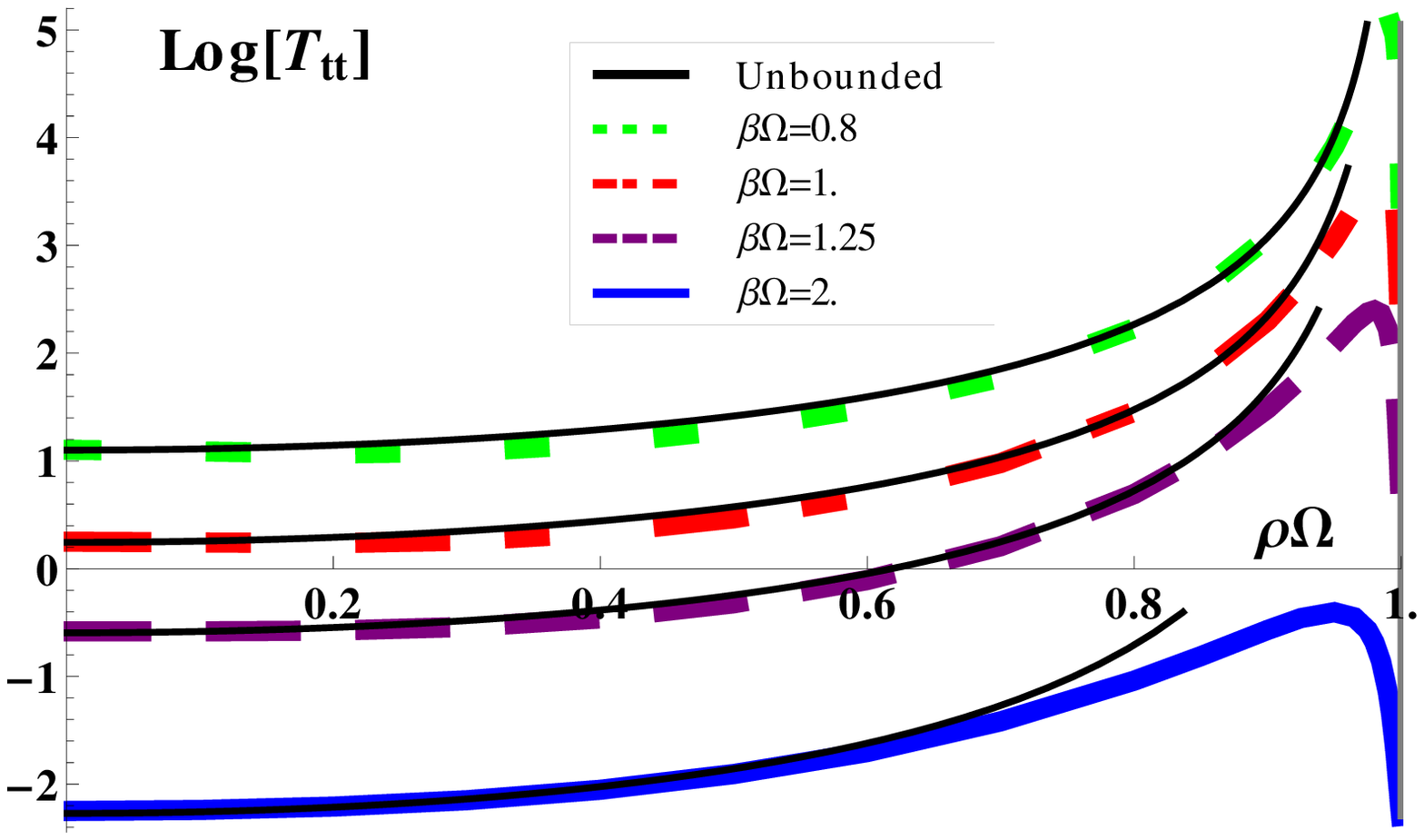}
    \figsubcap{b}
  \end{minipage}
    \caption{The t.e.v. of the energy density in the (a) spectral and (b) MIT bag
models for thermal fermions inside a boundary of radius $R=\Omega^{-1}$ is
compared with the corresponding energy density for the unbounded space-time
(thin curves), for four values of the inverse temperature $\beta$
($\beta$ is increasing from top to bottom).
The two models give different results near the boundary: in the spectral case, 
$T_{tt}$ decreases to a local minimum, then increases to a finite value on the
SOL, while in the MIT bag model, $T_{tt}$ increases to a local maximum, then
decreases to a finite value on the SOL which is more than two times greater
than in the spectral case.}
    \label{ave:plots}
\end{figure}
\section{Rotating fermions inside an infinite cylinder}\label{ave:cyl}
The divergent nature of expectation values in the rigidly rotating space-time
is related to the inclusion of a region of space which co-rotates at speeds
larger than the speed of light. Following Ref.~\cite{ave:duffy_ottewill},
we eliminate this region by enclosing the system inside a cylindrical boundary
of radius $R$, centred on the rotation axis, such that $R\Omega \le 1$. We use
two sets of boundary conditions, described below.

For consistency, a set of boundary conditions must
ensure the vanishing of the time derivative of the Dirac inner product,
given by:
\begin{equation}\label{ave:dtinner}
 i\partial_t \left<\psi, \chi\right> = \int_{\partial V} d \Sigma\,
 \overline{\psi} n_\mu \gamma^\mu \chi,
\end{equation}
where $n = n_\mu dx^\mu$ is the normal to the boundary.
In the spectral model \cite{ave:spectral}, Eq.~\eqref{ave:dtinner} is satisfied
mode by mode by Fourier coefficients of the wave functions $\psi$ and $\chi$.
In order to preserve the charge conjugation invariance of the theory, the
boundary conditions in the spectral model depend on the spectral index of the
mode, making this formulation non-local.
By contrast, the MIT bag model~\cite{ave:mit} satisfies Eq.~\eqref{ave:dtinner} in a
purely local way, by requiring that the wave functions satisfy the equation
$i n_\mu \gamma^\mu \psi = \psi$.
Plots of the t.e.v. of the energy density in the two models are
presented in Fig.~\ref{ave:plots}.
\section{The Casimir divergence}\label{ave:cas}
The presence of a boundary in a quantum system can induce a change in its
vacuum state.
The difference between the expectation value of the energy density for fermions of mass $\mu $ in
the vacuum states of the bounded and unbounded systems is found to
diverge as an inverse power of the distance to the boundary:
\begin{equation}
 \left<T_{tt}\right>_{\text{Casimir}} \sim
 \begin{cases}
  \displaystyle -\frac{1}{120\pi^2R^4} \times
  \frac{1 + 10\mu R}{(1 - \rho / R)^3} & \text{(MIT bag model)}\\
  \displaystyle -\frac{1}{16\pi^2R^4} \times
  \frac{1}{(1 - \rho / R)^4} & \text{(spectral boundary conditions.)}
 \end{cases}
\end{equation}

According to Ref.~\cite{ave:deutsch_candelas_boundary}, the energy density
should diverge as the inverse cube of the distance to the boundary for a purely
local stress-energy tensor. Although it seems to violate this result,
the spectral model gives a different order for the divergence because it does
not satisfy the assumption of locality.
\section*{Acknowledgments}
This work is supported by the Lancaster-Manchester-Sheffield Consortium for Fundamental Physics under STFC grant ST/J000418/1,
the School of Mathematics and Statistics at the University of Sheffield 
and European Cooperation in Science and Technology (COST) action MP0905 ``Black Holes in a Violent Universe''.

\bibliographystyle{plain}
\bibliography{main}

\begin{thebibliography}{1}

\bibitem{ave:rotfer}
Victor~E. Ambru\c{s} and Elizabeth Winstanley.
\newblock Rotating fermions.
\newblock {\em paper in preparation}.

\bibitem{ave:mit}
A.~Chodos, R.L. Jaffe, K.~Johnson, C.B. Thorn, and V.F. Weisskopf.
\newblock {N}ew extended model of hadrons.
\newblock {\em Phys. Rev. D}, 9(12):3471--3495, 1974.

\bibitem{ave:deutsch_candelas_boundary}
D.~Deutsch and P.~Candelas.
\newblock Boundary effects in quantum field theory.
\newblock {\em Phys. Rev. D}, 20(12):3063--3080, 1979.

\bibitem{ave:duffy_ottewill}
G.~Duffy and A.C. Ottewill.
\newblock Rotating quantum thermal distribution.
\newblock {\em Phys. Rev. D}, 67(044002), 2003.

\bibitem{ave:spectral}
M.~Hortac\c{s}u, K.D. Rothe, and B.~Schroer.
\newblock Zero-energy eigenstates for the {D}irac boundary problem.
\newblock {\em Nucl. Phys.}, B171:530--542, 1980.

\bibitem{ave:iyer}
B.~Iyer.
\newblock {D}irac field theory in rotating coordinates.
\newblock {\em Phys. Rev. D}, 26(8):1900--1905, 1982.

\bibitem{ave:letaw_pfautsch}
J.R. Letaw and J.D. Pfautsch.
\newblock {Q}uantized scalar field in rotating coordinates.
\newblock {\em Phys. Rev. D}, 22(6):1345--1351, 1980.

\bibitem{ave:vilenkin}
A.~Vilenkin.
\newblock {Q}uantum field theory at finite temperature in a rotating system.
\newblock {\em Phys. Rev. D}, 21(8):2260--2269, 1980.

\end{thebibliography}

\end{document}